\documentclass[prd,preprint,showpacs,preprintnumbers,amsmath,amssymb,floatfix]{revtex4}
\usepackage{amssymb}
\usepackage{graphicx}
\usepackage[export]{adjustbox}
\usepackage{bm}
\usepackage{mathrsfs}
\usepackage{verbatim}
\usepackage{wasysym}
\usepackage{slashed}
\usepackage{shellesc}
\usepackage[compat=1.1.0]{tikz-feynman}

\def\hf{{\frac{1}{2}}}

\begin{document}
\title{Quantization of Fields by Averaging Classical Evolution Equations}
\author{Timothy D. Andersen}
\affiliation{Georgia Institute of Technology, Atlanta, GA}
\email{andert@gatech.edu}
\date{\today}


\begin{abstract}
This paper extends the formalism for quantizing field theories via a microcanonical quantum field theory and Hamilton's principle to classical evolution equations. These are based on the well-known correspondence under a Wick rotation between quantum field theories and 4-D statistical mechanical theories. By placing quantum field theories on a 4+1-D under Wick rotation to 5-D, expectations of observables are calculated for a microcanonical field theory averaging Hamiltonian flow over a fifth spacelike dimension, a technique common in lattice gauge simulations but not in perturbation theory. In a novel demonstration, averaging pairs of external lines in the classical Feynman diagrams over the fifth dimension generates diagrams with loops and vacuum fluctuations identical to Standard Model diagrams. Because it is microcanonical, this approach, while equivalent for standard quantum fields theories in the Standard Model, is able to quantize theories that have no canonical quantization. It is also unique in representing expectations as averages over solutions to an ordinary, classical PDE rather than a path integral or operator based approaches. Hence, this approach draws a clear connection between quantum field theory and classical field theory in higher dimensions which has implications towards how quantum effects are interpreted. In particular, it raises questions about how violations of the ergodic hypothesis could influence quantum measurements even in standard, non-statistical quantum field theory.
\end{abstract}

\maketitle


\section{Introduction}
\label{sec:intro}
The correspondence between quantum field theory and classical equilibrium statistical mechanics has been well-established for decades \cite{McCoy:1994}, grounded in the path integral approach to quantum field theory \cite{Feynman:2010}. One way quantum mechanics differs from classical mechanics is that the latter obeys the principle of ``least" action (least meaning an extremum with first order change of zero) while the former obeys a statistical principle above least action where how close paths come to least action in a statistical rather than absolute sense depends on the value of Plank's reduced constant $\hbar$. The principle of least action states that the action $S$ is a time integral which is least for the motion between initial position and final and is represented as solving the extrema of $S$, typically by the variational principle, $\delta S = 0$. Quantum mechanics suggests that this is only approximately true in a statistical sense and that the action actually takes on many values above the least value. The same is true of statistical ensembles that exchange energy with a heat bath, and mathematically this equivalence emerges from the path integral. The path integral approach of Feynman obeys the complex distribution, ${Z}^{-1}e^{iS[\phi]/\hbar}$ with $Z = \int D\phi e^{iS[\phi]/\hbar}$ and $\phi$ a time-dependent field which approaches the least action principal as $\hbar\rightarrow 0$ \cite{Feynman:1948}. When time is taken to be imaginary (a Wick rotation \cite{Wick:1954}) $t\rightarrow i\tau$, the distribution becomes an ordinary Boltzmann distribution, ${Z}^{-1}e^{-\beta S}$ with $\beta = 1/(k_B T) = 1/\hbar$ with $T$, temperature, and $k_B$, Boltzmann's constant. Hence, $\hbar \rightarrow k_B T$ in Wick rotating. This correspondence can be extended from mechanics to field theory by adding spatial dimensions to the time dimension. Hence, there is a direct mathematical equivalence between quantum field theory and 4-D statistical mechanics of fields.

As discussed in the next few paragraphs, this correspondence is no mere analogy but a mathematical tool used frequently to go back and forth between classical and quantum computational and mathematical frameworks. Thus the equivalence is not only between quantum field theory and so-called canonical statistical mechanics. (The word ``canonical" means different things in different subfields. In statistical mechanics it means using Boltzmann distributions. In quantum field theory it means using operators rather than path integrals.) The correspondence extends mathematically from canonical statistical mechanics to microcanonical statistical mechanics by well-known equivalences \cite{Strominger:1983} in continuum field theories where the thermodynamic limit of particles or lattice points, $N\rightarrow\infty$, exists,
\[
{Z}^{-1}e^{-\beta S} = \Omega^{-1}\delta(A - S),
\] where $\Omega = \int D\phi \delta(A-S)$ and $\delta$ is the Dirac delta function. (For a proof, see Pauli's work \cite{Pauli:1973}.)

From microcanonical statistical mechanics, it extends to molecular dynamics \cite{Callaway:1983}, again by well-known equivalences that connect integrals of actions over configuration space with integrals of solutions to dynamical equations over infinite time. Establishing such a connection was necessary in the early 20th century in order to derive equilibrium statistical mechanics from first principles \cite{Pauli:1973}\cite{Gibbs:1902}. This connection depends on the assumption of ergodicity: the ability of a dynamical equation to cover all states in the configuration space over infinite time. Hence, with the assumptions of continuous fields, existence of a thermodynamic limit, and ergodicity a transitive connection can be made from canonical statistical mechanics to dynamics. These equivalences are represented by the following:

\begin{eqnarray}e^{iS/\hbar} \stackrel{\textnormal{Wick}}{\rightarrow} e^{-\beta S} \stackrel{\textnormal{TD limit}}{\equiv} \delta(A - S) \stackrel{\textnormal{ergodicity}}{\equiv}\nonumber\\ \left\{\frac{dp}{d\tau} = -\frac{\partial H}{\partial q}, \frac{dq}{d\tau} = \frac{\partial H}{\partial p}\right\} \stackrel{\textnormal{TD limit}}{\equiv}\nonumber \textnormal{Euler-Lagrange}
\end{eqnarray}

Given the equivalence under Wick rotation to equilibrium statistical mechanics in 4+1-D, all of these are likewise valid alternative approaches to quantum field theory.

Applications to quantum field theory emerged with the search for quantum gravity \cite{Strominger:1983b} and, separately, the expansion of computational lattice gauge theory. Microcanonical quantum field theory first appears in the literature in the early 1980's, initially as a contender for solving uncontrollable divergences in quantum gravity. Strominger developed a perturbation theory of the Standard Model in the microcanonical ensemble \cite{Strominger:1983}\cite{Strominger:1983b}. Fermions are addressed in his paper as well \cite{Strominger:1983} pp. 433-437 for the microcanonical ensemble. Strominger defines the action: $S = \int d^4x(\bar{\psi}\oslash^{-1}\psi + \bar{C}(x)\psi(x) + \bar{\psi}(x) C(x))$ where $\oslash^{-1}$ is a matrix operator, e.g., $\oslash^{-1} = \slashed{D} = \gamma^\mu (\partial_\mu - ieA_\mu)$ and $C(x)$ is some source field. The fields $\bar{\psi}(x)$ and $\psi(x)$ are anti-commuting spinor fields. Strominger's configuration integral is:
\[
\Omega[\bar{C},C,A] = \int D\psi_{F/2}D\bar{\psi}_{F/2} \delta(S - A),
\] where $A$ is the total fixed action and $F$ is the finite number of degrees of freedom defined by a box and ultraviolet cutoff. The argument to the delta function is an even order Grassmann algebra and Strominger computes the perturbation theory for the action.

In 1985, Iwazaki proved the equivalence of the full SU(N) lattice gauge theory with fermions in a Grassmann algebra and the standard functional equations under weak-coupling. In other words, the microcanonical ensemble gives the same perturbation series as the standard path integral \cite{Iwazaki:1985}. Both provide good predictions of weakly interacting fermions and bosons. At the end of his paper, he shows how, under ergodicity, the microcanonical formulation is equivalent to the time-average of Hamilton's equations, which leads to a molecular dynamical equivalence.

On the computational front, Creutz developed his demon-algorithm based microcanonical quantum field theory around the same time \cite{Creutz:1983}. Meanwhile, in a pair of seminal papers, Callaway developed microcanonical quantum field theory for lattice gauge computations \cite{Callaway:1982}\cite{Callaway:1983}. In his method, Callaway proposes that discrete fields such as the electromagnetic vector potential, $\phi_{n,\mu}$, have conjugate momenta with respect to a second time dimension $\tau$, $p_{n,\mu} = \partial \phi_{n,\mu}/\partial\tau$. He points out that in the canonical quantization any quantity independent of the field $\phi$ can be added to the action. Thus, an observable expectation with respect to an action $S$, \begin{equation}
\langle O\rangle = Z^{-1}\int D\phi O\exp[-S]
\label{eqn:canonical}
\end{equation} (let $\hbar=1$ henceforth), can also be given by $\langle O\rangle = Z'^{-1} \int D\phi\, Dp O\exp[-\beta H]$ where the energy functional, $H = T[p] + S[\phi]/\beta$, and partition function, $Z' = \int D\phi Dp \exp[-\beta H]$. He connects this equivalent theory to the microcanonical ensemble. Thus, Callaway was able to treat a quantum gauge theory as a classical dynamical theory in an additional dimension. These methods have since evolved into the hybrid Monte Carlo and molecular dynamics approaches to lattice gauge simulation \cite{Schroers:2001}. 

Approaching the problem as Hamiltonian flow in an additional dimension of $1...N$ lattice points of a 4-D lattice, the microcanonical quantum field theory is the integral over potential values on the lattice and their conjugate momenta:

\[
\Omega = \int d\phi_{1,\mu}\cdots d\phi_{N,\mu} dp_{1,\mu}\cdots dp_{N,\mu}\, \delta(E - H),
\] where the Hamiltonian is $H = T + V$ ($L=T-V$ is the Lagrangian by a Legendre transform), $T=\hf \sum_{n=1}^N |p_{n,\mu}|^2$ and $V=\sum \mathrm{Re }(1 - U_{n,\mu}U_{n+\mu}U^{-1}_{n+\nu,\mu}U^{-1}_{n,\nu})$ where $U_{n,\mu}=\exp i\phi_{n,\mu}$. The action is $S=\beta V$ where $\beta=1/g_0^2$ with $g_0$ a coupling constant. This formulation is for $U(1)$ lattice gauge theories but extends to SU(2) and SU(3).

Observables on the lattice are given by,
\begin{equation}
\langle O\rangle = \Omega^{-1}\int d\phi_{1,\mu}\cdots d\phi_{N,\mu} dp_{1,\mu}\cdots dp_{N,\mu}\, O\,\delta(E - H).
\label{eqn:O1}
\end{equation} The Hamiltonian is invariant under local gauge transformations $U_{n,\mu} + W_nU_{n,\mu} W_{n+\mu}^\dagger$ if $W_n\in U(1)$ and independent of $\tau$.

By Hamilton's equation for the flow through the second time dimensional $\tau$,
\[
\frac{d^2\phi_{n,\mu}}{d\tau^2} = \dot{p}_{n,\mu} = -\frac{\partial V\{\phi\}}{\partial \phi_{n,\mu}}.
\]

The flow explores the $(2N-1)$ dimensional hypersurface of constant energy $H[\phi,p]=E$ and any expectation value is given by the average,
\begin{equation}
\langle O\rangle = \lim_{\tau\rightarrow\infty} \frac{1}{\tau}\int_0^\tau d\tau'\, O[\phi(\tau'),p(\tau')],
\label{eqn:O2}
\end{equation} and is a unique function of $E$.

The equivalence of equations $\ref{eqn:O1}$ and $\ref{eqn:O2}$ follows from what Callaway refers to as the ``principle of equal weight'' which states that given $E$, trajectories given by solutions to Hamilton's equations cover the fixed energy hypersurface with equal density. That expectation values can be computed in either ensemble is discussed in \cite{Callaway:1982} and references cited therein. This principle follows from the ergodic hypothesis. As long as the initial conditions are chosen appropriately and the system has sufficient mixing, all points on the energy hypersurface will be visited with equal probability over infinite $\tau$. In this case, a time average is equivalent to an average over all phase space.

Hamilton's principle can then be applied via the Euler-Lagrange equations in the standard field theoretic way,
\begin{equation}
\frac{d}{d\tau}\left(\frac{\partial L}{\partial p}\right) + \frac{d}{dx_\mu}\left(\frac{\partial L}{\partial (\partial_\mu\phi)}\right) - \frac{\partial L}{\partial \phi} = 0,
\label{eqn:euler}
\end{equation} where $L$ is the Lagrangian $L=T-V$. In this case, we derive a partial differential equation indexed by coordinates $x$ rather than a set of ordinary differential equations indexed by lattice sites $n$. Provided that $L$ is not directly dependent on $\tau$, energy $H=E$ is conserved by these equations. This provides a starting point for a perturbation theory or a computational solution to the PDE.

Quantum field theory relies heavily on standard perturbation theory in the form of expanding non-linear parts of functionals in powers of weak coupling constants. This perturbation approach has been enormously successful, particularly for Quantum Electrodynamics \cite{Feynman:1948}, but has some well-noted problems as well. In classical evolution equations, large sections of the physical behavior may be inaccessible to standard perturbative expansions such as the strongly coupled phase of the Kardar–Parisi–Zhang (KPZ) above two dimensions \cite{Remez:2018}\cite{Wiese:1998}. Although not all power series fail to converge, divergent power series are inherently problematic in both statistical and dynamical physics, particularly for strong coupling. There are a number of alternative pertrubation methods such as Self-Consistent Expansion (SCE) which avoids the divergences of the standard perturbation theory for the KPZ equation. Variational perturbation theory can also convert divergent perturbation expansions in QFT into convergent ones \cite{Kleinert:1993} and has been applied to strong coupling in scalar field theory \cite{Kleinert:1998}.

Despite these drawbacks to perturbation theory, they underlie the primary tool for understanding fundamental particle interactions, Feynman diagrams. In this paper, I develop a perturbation theory in Feynman diagrams of the scalar field theory with quartic interaction, showing how they develop loops and vacuum contributions once they are integrated over ``time''. I address the equivalence of the perturbation theory on a classical 4+1-D evolution equation to the standard path integral-based perturbation theory of quantum field theory in the quartic interaction. I show how classical Feynman diagrams because quantum diagrams under ``time'' averaging, but I do not address the fundamental problem of asymptotic divergence. The efficacy of this theory to strongly interacting fields can be potentially addressed, however, with known methods for dealing with divergent perturbation series including those in QFT such as the variational and those in the study of classical evolution equations as in the SCE.

\section{Scalar Theory}
\label{sec:method}
We begin by assuming that all field theories exist on a 4+1-D flat manifold, such that the usual spacelike coordinates, $x,y,z$, are joined by a fourth spacelike dimension $w$ and the metric has signature $(-++++)$. We could take $w$ to be a timelike dimension as well or follow \cite{Overduin:1997} and leave it undetermined. In the case of a scalar theory it makes little difference, but, for other theories of interest, particularly gravity, this may create problems such as negative cosmological constants. More immediately, it is easier to present fermions on a de Sitter rather than anti-de Sitter spacetime \cite{Dirac:1935}. We use capital letters $A,B,C$ as 5-vector and 5-tensor subscripts, numbered 0 through 4 with $x_4=w$. Small Greek letters, $\mu,\nu,\lambda$, represent 4-vector and 4-tensors, numbered 0 through 3. All fields, $\psi,\phi,A_B,g_{AB},\dots$, are parameterized by this space-like dimension, $w$.  At every slice $w$ is a single instance or microstate of the classical 3+1-D universe. All vectors and tensors are assumed to have indexes from $0$ through $4$ (five indexes). We show below how assumptions on initial conditions, choice of gauge, and averaging can remove the additional index. Under Wick rotation $t\rightarrow i\sigma$, all fields are {\em in equilibrium} in the $w$ dimension; hence, actions are integrals over the four usual dimensions, $x_\mu = (t,x,y,z)$, which we write as $S=\int d^4x\mathcal{L}$. Also, let $\hbar=1$ for the following to avoid cluttering notation.

We will  work in double Wick rotated space to put the equations in more familiar statistical territory: $t\rightarrow i\sigma$ and $w\rightarrow i\tau$, making $\tau$ timelike and $\sigma$ spacelike. This has the benefit that reversing the Wick rotation of time gives us quantum field theoretic results while having the additional dimension timelike makes the calculations look more natural (wave rather than Poisson equations) while not affecting the results. This means the signature becomes $(++++-)$.

Given any standard action, $S$, over a field $\phi$ and coupling $g_0$ the corresponding Hamiltonian is the kinetic added to a potential energy, $V$,
\[
H = T + V,
\] where $S=\beta V$, $\beta=1/g_0^2$, and $T=\hf \int d^4x |p|^2$ where $p=\partial\phi/\partial \tau$. By a Legendre transform, $L = \left[\int d^4x |p\dot{x}|\right] - H$, we also obtain a Lagrangian,
\[
L = T - V.
\] If the field is a vector $A_\mu$, then $T=\hf \int d^4x p_\mu p^\mu$ and likewise for higher order tensors.

The $\phi^4$ Lagrangian is well-known \[
\mathcal{L}[\phi] = \hf\partial^\mu\phi\partial_\mu \phi + \hf m^2\phi^2 + \frac{\lambda}{4!} \phi^4 - J\phi,
\] Let $p=\partial \phi/\partial \tau$ in which case $L = \int d^4x \hf p^2 + \mathcal{L}[\phi]$. By the Euler-Lagrange equations, where $\ddot{\phi} = \partial^2\phi/\partial\tau^2$, the final evolution equation in terms of $\phi$ alone over the 4+1-D space is then the Klein-Gordon equation:
\begin{equation}
\ddot{\phi} = (\partial^\alpha\partial_\alpha - m^2)\phi - \frac{\lambda}{3!} \phi^3 + J.
\label{eqn:phi4}
\end{equation} or
\begin{equation}
0 = (\square_{4+1} + m^2)\phi + \frac{\lambda}{3!} \phi^3 - J
\label{eqn:phi4square},
\end{equation} where $\square_{4+1} = \partial^2_\tau - \partial^2_x - \partial^2_y - \partial^2_z - \partial^2_\sigma$.

\section{Perturbation Theory}

In this section, the derivation the perturbation theory is given for the scalar field. We will focus on the quartic theory, but the theory can be applied to other potentials.
\begin{equation}
0 = (\square_{4+1} + m^2)\phi + \frac{\lambda}{3!} \phi^3.
\label{eqn:phi4square2}
\end{equation}

We are interested in monomial potentials. Let $u(x,\tau)$ be a nonlinear potential such that our KG equation is,
\begin{equation}
0 = (\square_{4+1} + m^2)\phi + u(x,\tau).
\label{eqn:phi4square3}
\end{equation} Applying Green's function method to solve the nonhomogeneous equation,
\begin{equation}
\phi(x,\tau) = \int d^4y\,\int d\chi\, u(y,\chi)  D(x-y;\tau-\chi),
\label{eqn:phi0}
\end{equation} where the propagator (Green's function) is,
\begin{equation}
D(x-y;\tau-\chi) = \int \frac{d\omega}{(2\pi)} \frac{d^4k}{(2\pi)^4}\frac{e^{i(\omega(\tau-\chi) + (x-y)k)}}{\omega^2 - k^2 - m^2 + i\epsilon}
\label{eqn:kprop}
\end{equation} and $k^2= k_0^2 + k_1^2 + k_2^2 + k_3^2$, and $kx=k_\mu x^\mu$.

Now, we want to solve \ref{eqn:phi4square2} when $0<\lambda\ll m^2$. This method is straightforward perturbation theory \cite{Holmes:2012}. Let the solution be the perturbation series in $\lambda$,
\[
\phi = \sum_{n=0}^\infty \phi_{n} \lambda^n.
\] such that $\phi_{n}$ does not depend on $\lambda$. When we plug this into the equation we get,
\begin{align*}
& &\sum_{n=0}^\infty (\square_{4+1} + m^2)\phi_n\lambda^n = -\lambda\left(\sum_{n=0}^\infty \phi_n\lambda^n\right)^3 \\& = & -\sum_{n=0}^\infty \left(\sum_{\substack {k,l,m\\k+l+m+1=n}}\phi_k\phi_l\phi_m\right)\lambda^n.
\end{align*} Collecting coefficients we find,
\begin{equation}
(\square_{4+1} + m^2)\phi_n = -\sum_{\substack {k,l,m\\k+l+m+1=n}}\phi_k\phi_l\phi_m
\label{eqn:evoleqn}
\end{equation} for $n>0$ and 
\[
(\square_{4+1} + m^2)\phi_0 = 0.
\]

Because $k,l,m<n$, this creates an iterative solution. Because $k,l,m<n$, this creates an iterative solution. Starting with the free solution,
\[
(\square_{4+1} + m^2)\phi_0 = 0,
\] or, in momentum space,
\[
(-\omega^2 + k^2 + m^2)\hat{\phi}_0 = 0,
\] given by a sum of plane waves:
\begin{equation}
\hat{\phi}_0(k,\tau) = \frac{1}{2\omega(k)} \left\{A(k)e^{i\omega(k)\tau} + A^*(k)e^{-i\omega(k)\tau}\right\},
\label{eqn:planewave}
\end{equation} where $A(k)$ is any function, $\hat{\cdot} = \cal{F}[\cdot]$ is the Fourier transform $x_\mu\rightarrow k_\mu$, and $\omega(k) = \sqrt{k^2 + m^2}$. The position space free solution is $\phi_0(x,\tau) = \int d^4k/(2\pi)^4 e^{ikx}\hat{\phi}_0(k,\tau)$. We then use that solution to compute the next solution,
\[
(\square_{4+1} + m^2)\phi_1 = -\phi_0^3,
\] which is
\[
\phi_1(x,\tau) = \int d^4y\int d\chi \phi_0^3(y,\chi)D(x-y;\tau-\chi),	
\]then that for the following solution,
\[
(\square_{4+1} + m^2)\phi_2 = -3\phi_0^2\phi_1.
\] which allows our previous solution to substitute for $\phi_1$,
\begin{align*}
	\phi_2 &=& 3\int d^4y\int d\chi\int d^4y'\int d\chi'  \\ & &D(x-y;\tau-\chi)\phi_0^2(y,\chi)D(y-y';\chi-\chi')\phi_0^3(y',\chi'),
\end{align*} and so on so that all solutions can be found in terms of interactions of the free solution $\phi_0$ (\ref{eqn:phi0}). The integrals over $\chi$ and $\chi'$ indicate interactions between different slices of $\tau$, the classical 3+1-D universes.

Let the perturbation solution be truncated to level $N$, such that
\[
\phi(x,\tau) \approx \sum_{n=0}^N \phi_n\lambda^n,
\] and discard all solutions of order $O(\lambda^{n+1})$. (In general, the solution will diverge for some $\lambda>\epsilon\geq0$ as $N\rightarrow\infty$ as in standard perturbation theory.)

The expected value of the correlation of two plane wave solutions averaged over $\tau$ is,
\begin{eqnarray}
G_0(k,k')  & = & \langle \hat{\phi}_0(k)\hat{\phi}_0(k')\rangle \nonumber\\& = & \lim_{\tau\rightarrow\infty}\frac{1}{2\tau}\int_{-\tau}^{\tau}d\tau' \hat{\phi}_0(k,\tau')\hat{\phi}_0(k',\tau')\nonumber\\
& = & \lim_{\tau\rightarrow\infty}\frac{1}{2\tau}\int_{-\tau}^{\tau}d\tau'\frac{1}{4\sqrt{k^2+m^2}\sqrt{k'^2+m^2}} \nonumber\\ & \times &  \left\{A(k)A(k')e^{i(\omega(k)+\omega(k'))\tau'}\right. \nonumber\\
& + & A(k)A^*(k')e^{i(\omega(k)-\omega(k'))\tau'}\nonumber\\
& + & A^*(k)A(k')e^{i(-\omega(k)+\omega(k'))\tau'}\nonumber\\
& + & \left.A^*(k)A^*(k')e^{i(-\omega(k)-\omega(k'))\tau'}\right\}
\end{eqnarray} Now, note that the integral is only non-zero if $\omega(k)=\pm\omega(k')$. Since $\omega(k)\geq 0$, we have
\begin{eqnarray}
\langle \bar{\phi}_0(k)\bar{\phi}_0(k)\rangle & = & \lim_{\tau\rightarrow\infty}\frac{1}{2\tau}\int_{-\tau}^{\tau}d\tau' \phi(x,\tau')\phi(y,\tau')\nonumber\\ & = & \frac{|A(k)|^2}{2|k^2 + m^2|}.
\label{eqn:phi0corr}
\end{eqnarray} Let $A(k) = \sqrt{2}$, which gives the equivalent correlation to standard path integral quantization when $\hbar=c=1$. This gives the initial condition for $\phi$ as well (an amplitude scaling that turns out to be constant for all $k$). To regularize this for renormalization, which we will need to do, we should let $A(k)$ fall to zero at some large $|k|=\Lambda$, but for now we will let it remain constant.

We then use that solution to compute the next solution,
\[
(\square_{4+1} + m^2)\phi_1 = -\phi_0^3,
\] which is
\[
\phi_1(x,\tau) = \int d^4y\int d\chi \phi_0^3(y,\chi)D(x-y;\tau-\chi),	
\]then that for the following solution,
\[
(\square_{4+1} + m^2)\phi_2 = -3\phi_0^2\phi_1.
\] which allows our previous solution to substitute for $\phi_1$,
\begin{align*}
\phi_2 &=& 3\int d^4y\int d\chi\int d^4y'\int d\chi'  \\ & &D(x-y;\tau-\chi)\phi_0^2(y,\chi)D(y-y';\chi-\chi')\phi_0^3(y',\chi'),
\end{align*} and so on so that all solutions can be found in terms of interactions of the free solution $\phi_0$ (\ref{eqn:phi0}). The integrals over $\chi$ and $\chi'$ indicate interactions between different slices of $\tau$, the classical 3+1-D universes.

The $2M$-correlation Green's functions are,
\begin{align*}
G(x_1,\dots,x_{2M}) & = & \langle \phi(x_1)\cdots\phi(x_{2M})\rangle \\ & = & \lim_{\tau\rightarrow\infty} \frac{1}{\tau}\int_0^\tau d\tau' \phi(x_1,\tau)\cdots\phi(x_{2M},\tau).
\end{align*}

To compute these up to a truncated order $N$, we take the sum and compute all the cross terms in the $\tau$ integral up to the desired order $\lambda^N$, discarding higher order terms.

\section{Feynman Rules}
\label{sec:feynman}
Suppose we want the correlation up to order $\lambda$,
\begin{equation}
\langle \Omega| \phi(x_1)\phi(x_{2})|\Omega\rangle =  \lim_{\tau\rightarrow\infty} \frac{1}{\tau}\int_0^\tau d\tau' \phi(x_1,\tau)\phi(x_{2},\tau)\nonumber
\end{equation} Carrying out the perturbative calculation, the correlation has three terms. The first, a zeroth order correlation, involves two free fields and no interaction. The second and third involve correlation between a free field and a single interaction. 

These terms can be laboriously computed but there is a simpler approach with Feynman diagrams. Let the propagator be the field $\phi_g = D(x-y;\tau - \chi)$. The solution for position space $\phi_1(x,\tau)$, for example, is
\begin{center}
	\includegraphics{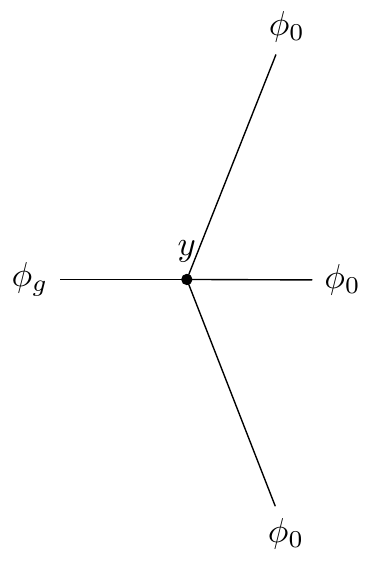}
\end{center}
This represents bringing three plane wave solutions, $\phi_0$, to a single point $y$ and transporting that to $x$ using the Green's function, $\phi_g$. This pattern can be iterated to higher orders adding more points that transport more particles together with symmetry factors representing the permutations of such graphs.

Helling \cite{Helling:2007} writes down the rules for these classical Feynman diagrams as follows:

\begin{enumerate}
	\item Draw n vertices for the expression for $\phi_n$ at order $\lambda^n$.
	\item Each vertex gets one in-going line at the left and three outgoing lines to the right.
	\item A line can either connect to the in-going port of another vertex or to the right-hand side
	of the diagram.
	\item Write down an integral for the point of each vertex.
	\item For a line connecting two vertices at points $y_1,\tau_1$ and $y_2,\tau_2$, write down a Green’s function $\phi_g(y_1 - y_2; \tau_1 - \tau_2)$.
	\item For a line ending on the right, write down a factor of $\phi_0$ evaluated at the point of the
	vertex at the left of the line.
	\item Multiply by the number of permutations of outgoing lines at the vertices which yield different diagrams (``symmetry factor'').
\end{enumerate}

These rules are nearly identical to the Feynman rules for the equivalent scalar statistical theory with the only difference being that loops are prohibited because of rules \#2 and \#3 on incoming and outgoing lines. Relaxing these rules allows for loops.

A loop in a Feynman diagram can represent a correlation of a point $y$ with itself, $\int_V d^4y \phi_g(y-y)$ or it can be two correlations of two points $y_1$ with $y_2$, $\int_V d^4y_1d^4y_2\phi_g(y_1 - y_2)\phi_g(y_2 - y_1)$, such that both points are integrated over a hypervolume $V$. (In momentum space, it can be one momenta $k$ integrated over a large box $\Lambda$.) Since these are integrated over all space and time, they form vacuum contributions in quantum field theory that affect the measurements of outcomes of experiments but are not directly measurable. 

While loops do not appear in the classical perturbation theory because the classical theory forms trees of diagrams, they do appear under averaging. The more orders of perturbations of the classical field theory the more branches, but there is no way for branches to double back and connect to prior branches. A plane wave at a point $y$, $\phi_0(y,\tau)$, however, can interact with another plane wave at $y$. These interactions, formed by simple multiplication of fields, are meaningless in the classical dynamical theory, but, integrated over infinite $\tau$, they are correlations in the statistical theory which then become loops. The way they become loops is that under averaging they are identical to 3+1-D propagators -- which is not the case in 4+1-D. (Recall: A propagator is simply a 2-point Green's function $G_0(x_1,x_2) = \phi_g(x_1-x_2) = \langle \phi_0(x_1)\phi_0(x_2)\rangle = \lim_{\tau\rightarrow\infty}\frac{1}{2\tau}\int_{-\tau}^\tau d\tau \phi_0(\tau,x_1)\phi_0(\tau,x_2)$.) Therefore, while $\int d^4y \phi_0(y,\tau)\phi_0(y,\tau)$ is just the product of two fields in 4+1-D, $\phi_g(y-y) = \lim_{\tau\rightarrow\infty}\frac{1}{2\tau}\int_{-\tau}^\tau d\tau \int d^4y \phi_0(y,\tau)\phi_0(y,\tau)$ is a vacuum loop at $y$ in 3+1-D \cite{Zee:2003}. For example, when the expected value is taken by averaging over $\tau$, we find that the correlation of pairs of plane wave solutions and the propagator, which are different in the 4+1-D evolution theory, become equal in the statistical theory. Reverse Wick rotating back to real time:
\begin{align*}
	\langle \phi_0(x)\phi_0(y)\rangle & = & \langle \phi_g(x-y)\rangle = \bar{\phi}_g(x-y) \\ & = & \int \frac{d^4k}{(2\pi)^4}\, \frac{e^{ik(x-y)}}{k^2 - m^2 + i\epsilon},
\end{align*} where $i\epsilon$ is a small value to avoid integrating through poles and $k^2 = -k_\mu k^\mu$. This can be shown by taking the $\tau$-average of \ref{eqn:kprop} and comparing it to the correlation given by \ref{eqn:phi0corr}. 

Expectations of order $\lambda$ correlations such as $\langle \phi_1(x_1)\phi_0(x_2)\rangle$, for example, develop loops in the 3+1-D statistical theory from the pairs of instances of $\phi_0$ that were external lines in the 4+1-D theory. The preceding example generates two pairs of plane wave solutions and one propagator which, when averaged, become three propagators. Because of the linearity of the integrals in the average over $\tau$, taking the expectation of correlations of perturbation solutions always reduces to products and sums of propagators provided the number of plane wave solutions is even.

In real time, a factor of $i$ is introduced in the example in the preceding paragraph: 
\begin{align}
&& G_1(x_1,x_2) = \langle \phi_1(x_1)\phi_0(x_2)\rangle\nonumber\\ & = & i\lambda \int d^4y\, \bar{\phi}_g(x_1 - y)\bar{\phi}_g(y-y)\bar{\phi}_g(x_2 - y)
\label{eqn:g1}
\end{align}. The diagrammatic equation correlates the first order diagram with a single, zeroth order plane wave solution (just a line),
\begin{widetext}
	\begin{equation}
	\left\langle \left( \includegraphics[valign=c]{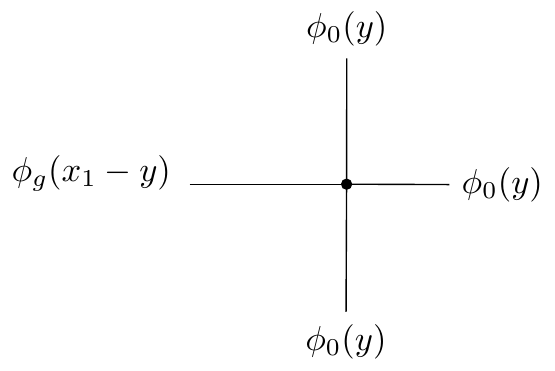} \right) \times 
	\includegraphics[valign=c]{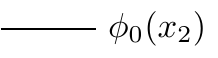}
	\right\rangle =
	\includegraphics{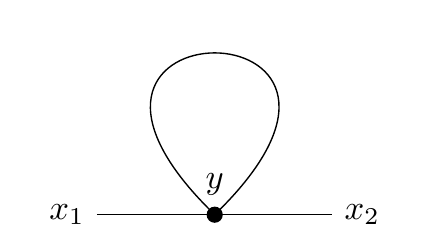}
	\end{equation}
\end{widetext} A pair correlation of plane waves at $y$ in the classical solution (the two solutions above and below the left diagram) becomes a propagator from $y$ back to itself in the quantum field solution on the right. Thus, it is as if two of the plane wave solutions in the diagram on the left connected to one another to form a propagator while the other plane wave solution at $y$ on the right side of the left diagram connects to the solution at $x_2$. 

\subsection{Momentum Space}
The $2M$-correlation Green's functions against the vacuum $\Omega$ in momentum space are,
\begin{align*}
G(k_1,\dots,k_{2M}) & = & \langle \Omega| \hat{\phi}(k_1)\cdots\hat{\phi}(k_{2M})|\Omega\rangle \\ & = & \lim_{\tau\rightarrow\infty} \frac{1}{\tau}\int_0^\tau d\tau' \hat{\phi}(k_1,\tau)\cdots\hat{\phi}(k_{2M},\tau).
\end{align*}

It can be obtained directly from the position space Green's function by Fourier transform,
\[
G(k_1,\dots,k_{2M}) = {\cal F}[G(x_1,\dots,x_{2M})];
\] Therefore, the same Feynman rules that apply in standard path integral quantum field theory apply in this case as well since the position space rules are the same.

The order $\lambda$ 2-point correlation in Minkowski space, for example, is the Fourier transform of \ref{eqn:g1},
\begin{align*}
G_1(k) & = & {\cal F}[G_1(x_1,x_2)] = \langle \phi_1(k,\tau)\phi_0(k,\tau)\rangle \\ & = & i\lambda\int \frac{d^4p}{(2\pi)^4} \frac{1}{p^2 - m^2 + i\epsilon}\left(\frac{1}{k^2 - m^2 + i\epsilon}\right)^2
\end{align*}

\section{Perturbation Series Equivalence}
The perturbation series can be shown to be equivalent to the standard path integral series which is often given the functional definition,
\begin{align*}
Z[J] & = & \sum_{n=0}^{\infty} \frac{1}{n!}\left(\frac{\lambda}{4!}\right)^n\times \\
& &\int d^4x_1\cdots d^4x_n \frac{\delta^4}{\delta J(x_1)}\cdots\frac{\delta^4}{\delta J(x_n)} Z_0[J],
\end{align*} where $Z_0[J]$ is the free path integral and $Z[J]$ the fully interacting one. This series, while divergent for strong coupling, is the basis for Feynman diagrams. Let the scalar action be $S = S_0 + \lambda S_{int}$, dividing it into the free part $S_0$ and the interacting part $S_{int}$, and the Hamiltonian is $H= H_0 + \lambda S_{int}$ where $H_0 = T + S_0$ (with $T = -\hf p^2 = -\hf \dot{\phi}^2$ for scalar theory in 4+1-D). Iwazaki proved the equivalence for the scalar theory for the microcanonical quantum field theory \cite{Iwazaki:1984},
\begin{align*}
\Omega & = & \int DpD\phi\, \delta(E - H_0 - \lambda S_{int}) \\ & = & \sum_{n=0}^\infty \frac{(-1)^n}{n!}\frac{d^n}{dE^n} \int DpD\phi \delta(E - H_0)(\lambda S_{int})^n
\end{align*} by convergence of the continuum limit on a lattice. Given an observable, $O[\phi]$ for which the ensemble is convergent, e.g., any Green's function, it has a series in $\lambda$,
\[
\langle O[\phi] \rangle = \sum_{n=0}^\infty \lambda^n\langle O_n[\phi]\rangle.
\] Thus, Iwazaki's result proves,
\begin{align*}
&&\sum_{n=0}^N \lambda^n\langle O_n[\phi]\rangle  \\ & = & \Omega_N^{-1}\sum_{n=0}^N \frac{(-1)^n}{n!}\frac{d^n}{dE^n} \int Dp D\phi O_n[\phi]\delta(E - H_0)(\lambda S_{int})^n\\
& = & Z_N^{-1}\sum_{n=0}^{N} \frac{(-1)^n}{n!} \int D\phi O_n[\phi]e^{-S_0}(\lambda S_{int})^n
\end{align*} 
for any $N$ given that $\Omega_N = \sum_{n=0}^N \frac{(-1)^n}{n!}\frac{d^n}{dE^n} \int Dp D\phi \delta(E - H_0)(\lambda S_{int})^n$ and $Z_N = \sum_{n=0}^N \frac{(-1)^n}{n!} \int D\phi e^{-S_0}(\lambda S_{int})^n$. Wick rotating to Minkowski, this shows the equivalence of the microcanonical and canonical path integral approaches to quantum field theory for scalar theory. This result has also been proved for SU(N) and fermions \cite{Iwazaki:1985}.

What remains to show is that the perturbation series of the evolution equation approach is equivalent to Iwazaki's microcanonical ensemble. Starting with the Hamiltonian in the double Wick rotated de Sitter spacetime metric,
\begin{equation}
H = \hf[-\dot{\phi}^2 + (\partial_\mu \phi)^2 + m^2\phi^2] + (\lambda/4!)\phi^4,
\label{eqn:h}
\end{equation} we set $E=H$ as the fixed energy of the 4+1-D system. We know that $E$ is a function of $\lambda$ and given the power series for small $\lambda$, $E = \sum_{n=0}^\infty E_n\lambda^n$. Further, we have the power series of $\phi = \sum_{n=0}^\infty \phi_n\lambda^n$. The Hamiltonian breaks into,
\begin{align*}
&&\sum_{n=0}^\infty E_n\lambda^n = \\ &&\hf\left[-(\sum_{n=0}^\infty \dot{\phi_n}\lambda^n)^2 + (\sum_{n=0}^\infty \partial_\mu\phi_n\lambda^n)^2 + m^2(\sum_{n=0}^\infty \phi_n\lambda^n)^2\right]\\ && + (\lambda/4!)(\sum_{n=0}^\infty \phi_n\lambda^n)^4.
\end{align*}
 This equation, by matching terms, generates an infinite number of equations that are iteratively solvable, i.e. equation $n+1$ can be solved with the solution to $n$,
\begin{align}
E_n  & = &  H_n = \hf\sum_{\substack{k,l\\k+l=n}} [-\dot{\phi}_k\dot{\phi}_l + \partial_\mu \phi_k \partial^\mu \phi_l + m^2\phi_k\phi_l] \nonumber\\ & + & (4!)^{-1}\sum_{\substack{j,k,l,m\\j+k+l+m+1=n}}\phi_j\phi_k\phi_l\phi_m.
\label{eqn:en}
\end{align}

Each Hamiltonian, by the application of Euler Lagrange generates an evolution equation above (\ref{eqn:evoleqn}). Taking the microcanonical statistical ensemble from the equation for $H$ (\ref{eqn:h}) and setting it equal to the statistical ensemble for the infinite set of equations for $H_n$, up to a constant ${\cal C}$, the equivalence is,
\begin{equation}
\Omega = \int D\phi \delta(E - H) = {\cal C}\prod_{n=0}^\infty \int D\phi_n \delta(E_n - H_n) = {\cal C}\Omega'.
\label{eqn:omega}
\end{equation} Since $\phi_{n+1}$ is determined by $\phi_n$, all the functions are determined by $\phi_0$. Likewise, $E_{n+1}$ is determined by $E_n$. Thus, up to a constant, the infinite number of functional integrals collapses to one functional integral over one field $\phi_0$ and one independent energy constant, $E_0$, which is determined by $\hbar$. Thus, we can show that for the observable $O$ that the microcanonical series given by Iwazaki is equivalent to the microcanonical series given by \ref{eqn:omega},
\begin{align}
	&& \sum_{n=0}^N \lambda^n\langle O_n[\phi]\rangle \nonumber\\ & = & \Omega_N^{-1}\sum_{n=0}^N \frac{(-1)^n}{n!}\frac{d^n}{dE^n} \int D\phi O_n[\phi]\delta(E - S_0)(\lambda S_{int})^n\nonumber\\
	& = & \Omega_N'^{-1}   \sum_{n=0}^{N} \lambda^n \int D\phi_0 O_n[\phi] \prod_{m=0}^n\delta(E_m - H_m),
	\label{eqn:result}
\end{align} neglecting orders of $\lambda$ higher than $N$. This final result follows because the coefficients to powers $\lambda^n$ must match on both sides since they determine the same expectations of the observable to the required order in the Taylor series. In other words, we have two derivations of the same perturbation series from the same original non-perturbed ensemble which necessarily determine the same expectations of any observable to the same order if that expectation exists (which it does for point-correlations for sufficiently small $\lambda$).

Assuming ergodicity, since the PDE \ref{eqn:en} is derived from the energy functional \ref{eqn:evoleqn} by Euler-Lagrange, the resulting expectations of observables from each must be equivalent. This observation along with \ref{eqn:result} proves that the time-averaging approach is equivalent for perturbation theory as it is for the non-perturbative ensemble and
\[
\sum_{n=0}^N \lambda^n\langle O_n[\phi]\rangle = \sum_{n=0}^N \lambda^n\lim_{\tau\rightarrow\infty}\frac{1}{2\tau}\int_{-\tau}^\tau d\tau'  O_n[\phi,\tau'],
\] gives the standard path integral result up to order $N$.

\section{Conclusion}
Discrete Hamiltonian flow method of molecular dynamics has been extended to a continuum PDE method showing that classical Feynman diagrams reduce to quantum field theoretic ones when correlated. By doing so, a novel way of reducing quantum physics on a 3+1-D spacetime to classical physics on a 4+1-D spacetime has been presented. Of interest are cases where ergodicity may break down strongly enough to be measureable. New experimental results would be required to determine if ergodicity is a factor in quantum expectations. A classic classical example of a system that violates ergodicity is the Fermi-Pasta-Ulam-Tsingou (FPUT) system involving an oscillating string with a quadratic non-linearity \cite{Fermi:1955}. In general, exactly integrable systems are the most likely to exhibit non-ergodic behavior since they are integrated from initial conditions and can be periodic rather than ``forgetting'' initial conditions. In this case, the predictions of the time-averaging approach will diverge from those of the path integral. This would provide a test to determine if Wick rotated quantum field theory differs from equilibrium statistical mechanics. 

In a strong breakdown of ergodicity, a statistical ensemble continues to obey the dynamical (evolution) equations but violates the path integral ensemble (or only obeys a subset of it) because the regions of microstates on the energy manifold covered are different, with the non-ergodic system avoiding trajectories because of its initial conditions. Phase transitions are another example where regions of microstates are avoided, not because of initial conditions but because of the parameters of the system such as inverse temperature which determine the phase. The key question here is to understand the relationship between quantum theory as a 3+1-D path integral and the classical counterpart in 4+1-D that are equivalent under the ergodic hypothesis and which is more fundamental. If violations to that hypothesis could be constructed experimentally through a quantum analog of the FPUT system, for example, the existence of an additional dimension would be implied. If not, then quantum theory would be shown to be fundamentally different from classical equilibrium statistical mechanics derived from dynamical first principles. Such a discovery would be important to the interpretation of quantum phenomena. Future work is to extend the ideas here to SU(N) and fermions and to address the problem of ergodicity in quantum field theory in order to give more insight into such ergodicity violating systems.

\end{document}